\documentclass[prd,showpacs,preprintnumbers,amsmath,amssymb,nofootinbib]{revtex4}

\usepackage{graphicx}
\usepackage{dcolumn}
\usepackage{bm}

\usepackage{color}
\def\red{\color{red}}

\begin{document}

\title{Dyon Production from Near-Extremal Kerr-Newman-(Anti-)de Sitter Black Holes}

\author{Chiang-Mei Chen} \email{cmchen@phy.ncu.edu.tw}
\affiliation{Department of Physics, National Central University, Chungli 32001, Taiwan}
\affiliation{Center for High Energy and High Field Physics (CHiP), National Central University, Chungli 32001, Taiwan}

\author{Sang Pyo Kim}\email{sangkim@kunsan.ac.kr}
\affiliation{Department of Physics, Kunsan National University, Kunsan 54150, Korea}

\date{\today}

\begin{abstract}
Using the enhanced symmetry in the near-horizon region of the near-extremal dyonic Kerr-Newman (KN) black hole in the (A)dS space, we find the exact solutions for dyonic charged scalar field in terms of the hypergeometric function and explicitly compute the Schwinger effect for the emission of electric and/or magnetic charges. The emission formula confirms a universal factorization of the Schwinger formula in the AdS$_2$ and another Schwinger formula in the two-dimensional Rindler space determined by the effective temperature and the Hawking temperature with the chemical potentials of electric and/or magnetic charges and the angular momentum. The emission of the same species of charges from the KN black hole is enhanced in the AdS boundary while it is suppressed in the dS boundary. In addition, the dragging of particles in the KN black hole diminishes the emission of charges in both AdS and dS spaces. The AdS geometry of near-horizon gives the Breitenloher-Freedman (BF) bound, within which the stability of dyonic KN black holes is guaranteed against both the emission of charges and Hawking radiation.
\end{abstract}


\maketitle

\section{Introduction}
The Einstein-Maxwell theory in four dimensions has Reissner-Nordstr\"{o}m (RN) black holes and Kerr-Newnman (KN) black holes in the asymptotically flat spacetime. The RN and KN black hole solutions are also known in the (anti-)de Sitter space (for review and references, see~\cite{Griffiths:2012}). In the asymptotically flat spacetime the RN and KN black holes have, in addition to the mass and angular momentum, one more hair (parameter), the charge of the hole, and show a richer structure than Schwarzschild and Kerr black holes. The RN black hole, for instance, has two horizons: the event horizon and the Cauchy (inner) horizon. When two horizons coincide and form a degenerate horizon, the RN black hole becomes an extremal black hole and is described by an ${\rm AdS}_2 \times {\rm S}^2$ geometry near the horizon~\cite{Kunduri:2008rs}, in which quantum fields can be analytically studied. The KN black hole has, in addition to mass $M$, two additional hairs (parameters), the angular momentum $a$ as well as the charge $Q$ and can become an extremal black hole in the limit of $M^2 = a^2 + Q^2$, whose near-horizon geometry is a warped ${\rm AdS}_3 \times {\rm S}^1$~\cite{Chen:2016caa}. In the (A)dS space the RN and KN black hole has an extra parameter, the radius $L$ of the (A)dS space. Hence, the RN and KN black holes in the (A)dS space exhibit much richer structure than those in the asymptotically flat space. For instance, the near-horizon geometry ${\rm AdS}_2 \times {\rm S}^2$ of the extremal RN black hole that is obtained by coinciding the event horizon with the Cauchy horizon has two different radii $R_{AdS}$ and $R_S$~\cite{Chen:2020mqs}. Furthermore, the RN black hole has the Nariai limit when the event horizon and the cosmological horizon coincide, whose near-horizon geometry is ${\rm dS}_2 \times {\rm S}^2$ (see, for instance,~\cite{Montero:2019ekk}.)

Hawking radiation exhibits part of quantum aspects of black holes~\cite{Hawking:1974sw}. A puzzling property of Hawking radiation is that the vanishingly small Hawking temperature for near-extremal black holes exponentially suppresses Hawking radiation. The black hole thermodynamics is not apparently obvious for extremal black holes. In the case of charged black holes in an asymptotically flat space, the electric field on the horizon is strong enough to discharge the hole's charge via the Schwinger mechanism~\cite{Sauter:1932ab, Schwinger:1951nm} and in the (A)dS space~\cite{Belgiorno:2007va, Belgiorno:2008mx}. Further, the quantum field of a massive charge in the near-extremal black holes can be solved in terms of known functions, and the Schwinger emission of charges has been explicitly found for the RN and KN black holes in the asymptotically flat space~\cite{Chen:2012zn, Chen:2014yfa, Chen:2016caa, Chen:2017mnm}. Recently, the Schwinger emission from (near-)extremal RN black holes also was found in the (A)dS space~\cite{Chen:2020mqs, Cai:2020trh, Zhang:2020apg}. The AdS boundary compresses the horizon radius, on which the electric field becomes stronger than that in the asymptotically flat space. Hence, the Schwinger effect for a fixed mass is enhanced in the AdS space. In contrast, the dS boundary stretches the horizon radius and the weakened electric field suppresses the Schwinger effect.

The rotation of near-extremal KN black hole affects the Schwinger effect via two opposite factors: decreasing the ``inertial mass'' of charge due to the spacetime dragging but increasing the horizon radius. Hence, the rotation of the black hole lowers the Unruh temperature for the charge acceleration on the horizon and thereby the effective temperature for the leading Boltzmann factor for the Schwinger effect~\cite{Chen:2016caa, Chen:2017mnm}. In fact, the spacetime dragging around the event horizon affects the radial equation for the charges reducing the ``inertial mass''.

In this paper, we study the Schwinger effect from the near-extremal KN black holes in the (A)dS space. This model is physically interesting since the rotation of the black hole, with constrained mass by near-extremal condition, reduces the Schwinger effect while the AdS (dS) boundary increases (decreases) the Schwinger effect. Thus, the rotation cooperates the suppression in the dS boundary but competes for the enhancement with the AdS boundary. We show that the Unruh temperature for charge acceleration and the effective temperature for Schwinger effect are higher in the AdS space than in the asymptotically flat space while they are lower in the dS space than in the flat space. We then extend the Schwinger effect to the emission of dyons with electric and magnetic charges and find a universal formula for the mean number of produced dyons from the near-extremal KN black holes in the (A)dS space.

The organization of this paper is as follows. In Sec.~\ref{secII}, we study the near-horizon geometry of near-extremal KN black holes in the (A)dS space. The degenerate horizon of extremal KN (A)dS black holes gives a constraint among the mass, angular momentum, charges and (A)dS radius, which expresses one parameter in terms of the others. Here we parameterize the degenerate horizon and charges by the mass, angular momentum and the radius of the (A)dS space, and also express the degenerate horizon and the mass in terms of the charges, angular momentum and the radius of the (A)dS space. In Sec.~\ref{secIII}, we analyze the quantum field for dyons in the near-horizon geometry and find the mean number of produced dyons for the Schwinger effect. We numerically study the effect of the angular momentum and the radius of the (A)dS space on the horizon radius, inertial mass, Unruh temperature as well as the effective temperature. The BF bound is found for the KN black holes in the AdS space. In Sec.~\ref{secIV}, we discuss the effect of the rotation and the (A)dS boundary on the Schwinger effect for emission of charges, electric and/or magnetic charges from the KN black holes.

\section{Extremal dyonic KN Black Holes in (A)dS Space}\label{secII}
\begin{figure}[ht]
\begin{center}
\includegraphics[width=0.8\textwidth]{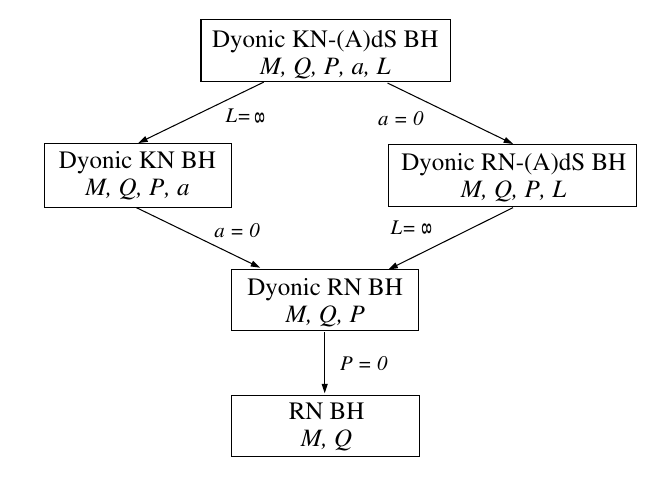}
\caption{Chart of charged black holes from the (A)dS space to the asymptotically flat space.} \label{fig_BH}
\end{center}
\end{figure}

In this section we express the near-horizon geometry of near-extremal KN black holes in the (A)dS space. The KN-(A)dS black holes are the most general stationary solutions, as shown in Fig.~\ref{fig_BH}, where the KN black hole in the asymptotically flat space and the RN black hole in the (A)dS space are the limit of the infinite (A)dS radius and the zero angular momentum, respectively. The four-dimensional dyonic KN-(A)dS black holes consisting of four physical quantities, the mass $M/\Xi^2$, angular momentum $J = M a/\Xi^2$, electric and magnetic charges $Q/\Xi$ and $P/\Xi$, have the metric and the gauge potential
\begin{eqnarray}
ds^2 &=& - \frac{\Delta_r}{\Sigma} \left( dt - \frac{a \sin^2\theta}{\Xi} d\varphi \right)^2 + \frac{\Sigma}{\Delta_r} dr^2 + \frac{\Sigma}{\Delta_\theta} d\theta^2
\nonumber\\
&+& \frac{\Delta_\theta}{\Sigma} \sin^2\theta \left( a dt - \frac{r^2 + a^2}{\Xi} d\varphi \right)^2,
\nonumber\\
A_{[1]} &=& \frac{Q r - P a \cos\theta}{\Sigma} \left( dt - \frac{a \sin^2\theta}{\Xi} d\varphi \right) + \frac{P (\cos\theta - \sigma)}{\Xi} d\varphi,
\end{eqnarray}
where
\begin{eqnarray}
&& \Delta_r = (r^2 + a^2) \left( 1 \pm \frac{r^2}{L^2} \right) - 2 M r + Q^2 + P^2, \qquad \Delta_\theta = 1 \mp \frac{a^2}{L^2} \cos^2\theta,
\nonumber\\
&& \Sigma = r^2 + a^2 \cos^2\theta, \qquad \Xi = 1 \mp \frac{a^2}{L^2}.
\label{param}
\end{eqnarray}
Here the upper (lower) sign corresponds to the AdS (dS) space and $\sigma = \pm 1$ denotes a different choice of the Dirac string.
The dual gauge potential is
\begin{equation}
\bar A_{[1]} = \frac{P r + Q a \cos\theta}{\Sigma} \left( dt - \frac{a \sin^2\theta}{\Xi} d\varphi \right) - \frac{Q (\cos\theta - \sigma)}{\Xi} d\varphi.
\end{equation}
The associated thermodynamic quantities, such as the Hawking temperature, entropy, horizon angular velocity and chemical potential are given by, respectively,
\begin{eqnarray} \label{TSOP}
&& \hat T_\mathrm{H} = \frac{r_+}{4 \pi (r_+^2 + a^2)} \left( 1 - \frac{a^2 + Q^2 + P^2}{r_+^2} \pm \frac{a^2}{L^2} \pm \frac{3 r_+^2}{L^2}  \right),
\nonumber\\
&& \hat S_\mathrm{BH} = \frac{\pi (r_+^2 + a^2)}{\Xi}, \qquad \hat \Omega_\mathrm{H} = \frac{a \Xi}{r_+^2 + a^2}, \qquad \hat \Phi_\mathrm{H} = \frac{Q r_+}{r_+^2 + a^2}. 
\end{eqnarray}

The extremal limit with a degenerate horizon at $r_0$ can be achieved by requiring the conditions $\Delta_r(r_0) = \Delta_r'(r_0) = 0$ that identify the relation of physical parameters. We introduce two convenient ways to express the physical quantities: first in terms of mass and then in terms of charges. In the dS space the degenerated horizon $r_0 = r_+ = r_-$ is given, in terms of mass, by
\begin{eqnarray} \label{r0_M-a-L1}
r_0 &=& \sqrt{\frac23} \left( 1 - \frac{a^2}{L^2} \right)^{1/2} L \sin\frac{\vartheta}{3},
\nonumber\\
Q^2 + P^2 &=& \frac23 \left( 1 - \frac{a^2}{L^2} \right)^2 L^2 \sin^2\frac{\vartheta}{3} \cos\frac{2 \vartheta}{3} - a^2,
\end{eqnarray}
where
\begin{equation}
\sin\vartheta = 3 \sqrt{\frac32} \left( 1 - \frac{a^2}{L^2} \right)^{-3/2} \frac{M_0}{L}, \qquad 0 \le \vartheta \le \frac{\pi}2.
\end{equation}
There is an upper limit for the mass and its maximum value corresponds to $\theta = \pi/2$. A simple way to give the associated physical quantities in AdS space is analytically continuing $L \to - iL, \vartheta \to i \zeta$, or equivalently flipping the sign of the cosmological constant, leading to
\begin{eqnarray} \label{r0_M-a-L2}
r_0 &=& \sqrt{\frac23} \left( 1 + \frac{a^2}{L^2} \right)^{1/2} L \sinh\frac{\zeta}{3},
\nonumber\\
Q^2 + P^2 &=& \frac23 \left( 1 + \frac{a^2}{L^2} \right)^2 L^2 \sinh^2\frac{\zeta}{3} \cosh\frac{2 \zeta}{3} - a^2,
\end{eqnarray}
where
\begin{equation}
\sinh\zeta = 3 \sqrt{\frac32} \left( 1 + \frac{a^2}{L^2} \right)^{-3/2} \frac{M_0}{L}.
\end{equation}
Another parametrization of the extremal KN black holes, in terms of charges, gives\footnote{There is another degenerate limit in the dS space, namely the Nariai branch $r_0 = r_+ = r_C$, in terms of mass,
\begin{eqnarray*}
r_0 &=& \sqrt{\frac23} \left( 1 - \frac{a^2}{L^2} \right)^{1/2} L \sin\left( \frac{\pi}3 - \frac{\vartheta}3 \right),
\\
Q^2 + P^2 &=& \frac23 \left( 1 - \frac{a^2}{L^2} \right)^2 L^2 \sin^2\left( \frac{\pi}3 - \frac{\vartheta}3 \right) \cos\left( \frac{2 \pi}3 - \frac{2 \vartheta}3 \right) - a^2,
\end{eqnarray*}
or in terms of charges,
$$ r_0^2 = \frac{L^2}6 \left( \sqrt{ \left( 1 - \frac{a^2}{L^2} \right)^2 - \frac{12 (a^2 + Q^2 + P^2)}{L^2}} + \left(1 - \frac{a^2}{L^2} \right) \right). $$
The corresponding near-horizon geometry becomes warped dS$_3$ since the signatures of $d\tau$ and $d\rho$ are flipped.}
\begin{equation} \label{r0_a-Q-L}
r_0 = \frac{L \sqrt{\delta}}{\sqrt6}, \qquad M_0 = \frac{L \left[ 3 \left( 1 \pm \frac{a^2}{L^2} \right) \pm \delta \right] \sqrt{\delta}}{3 \sqrt6},
\end{equation}
where
\begin{equation}
\delta = \pm \left( \sqrt{ \left( 1 \pm \frac{a^2}{L^2} \right)^2 \pm \frac{12 (a^2 + Q^2 + P^2)}{L^2}} - \left( 1 \pm \frac{a^2}{L^2} \right) \right).
\end{equation}

A few comments are in order. Extremal black holes require that $\Delta_r$ in~(\ref{param}) should have two equal roots, which lead to the zero discriminant for the quartic equation~\cite{Dickson:1914} and give a constraint equation among parameters:
\begin{eqnarray}
&& \left( \Bigl(1 \pm \frac{a^2}{L^2} \Bigr)^3 \pm 54 \frac{M_0^2}{L^2} \mp 36 \Bigl(1 \pm \frac{a^2}{L^2} \Bigr) \frac{Q^2+ P^2 + a^2}{L^2} \right)^2
\nonumber\\
&=& \left(\Bigl(1 \pm \frac{a^2}{L^2} \Bigr)^2 \pm 12  \frac{Q^2+ P^2 + a^2}{L^2}  \right)^3.
\label{discriminant}
\end{eqnarray}
A large $L$ expansion of~(\ref{discriminant}) gives $(M_0^2 - (Q^2 + P^2 + a^2))/L^2$ up to ${\cal O} (1/L^4)$ and recovers $M_0 = \sqrt{Q^2 + P^2 + a^2}$ in the asymptotically flat space ($L = \infty$).
Notice that the second equation in~(\ref{r0_M-a-L1}) and Eq.~(\ref{r0_a-Q-L}) satisfy the constraint~(\ref{discriminant}) while the first equation
in~(\ref{r0_M-a-L1}) and Eq.~(\ref{r0_a-Q-L}) determine the horizon in terms of $(M_0, a, L)$ and $(a, Q^2+P^2, L)$, respectively. Still another parametrization was introduced in Ref.~\cite{Hartman:2008pb} that expresses the horizon radius as a function of $M_0, Q^2 + P^2$ and $L$.

\begin{figure}[ht]
\begin{center}
\includegraphics[width=0.8\textwidth]{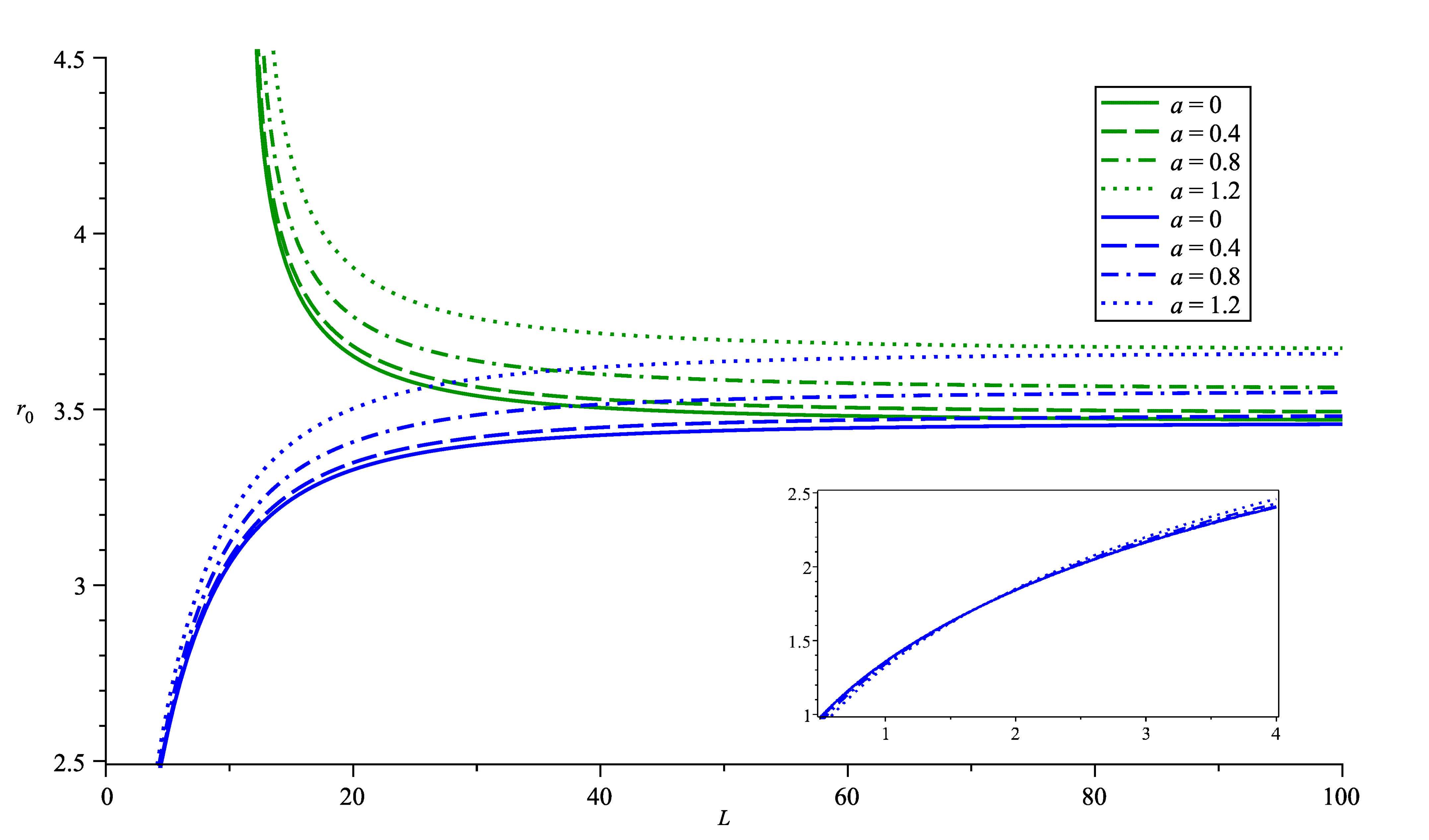}
\caption{Horizon radius $r_0$ in AdS (blue) and dS (green) with parameters $Q^2 + P^2 = 12$. The dS space increases the horizon radius while the AdS space decreases the radius. But the horizon radius approaches the flat space limit when the radius of the (A)dS space becomes large. The rotation increases the horizon radius both in the (A)dS space.} \label{fig_r0}
\end{center}
\end{figure}

As discussed in~\cite{Chen:2020mqs}, the boundary AdS space decreases the horizon radius while the dS space increases the radius. For an extremal rotating charged BH, the rotation provides an additional repulsive centrifugal force for produced particles requiring, with fixed charge, larger mass which increases the horizon radius in both dS and AdS spaces, see Fig.~\ref{fig_r0}. Moreover, in AdS space, the contributions from $a$ and $L$ are competitive so that their dominance switches at almost the same value of $r_0$ and $L$ corresponding to the U-turn (convexity) in the Unruth temperature.

The near-horizon geometry of $r_+ = r_- = r_0$ can be obtained by applying the following transformation
\begin{equation}
\varphi \to \varphi + \hat \Omega_\mathrm{H} \, t, \quad r \to r_0 + \epsilon \, \rho, \quad t \to \frac{r_0^2 + a^2}{\Delta_0} \, \frac{\tau}{\epsilon}, \quad M \to M_0 + \frac{\Delta_0}{2 r_0} B^2 \epsilon^2,
\end{equation}
and then taking $\epsilon \to 0$ limit. The parameter $B$ characterizes the deviation from the extremal limit. Finally, the near-horizon geometry of KN-(A)dS black hole is given by
\begin{eqnarray}
ds^2 &=& \frac{\Sigma_0}{\Delta_0} \left( - (\rho^2 - B^2) d\tau^2 + \frac{d\rho^2}{\rho^2 - B^2} + \frac{\Delta_0}{\Delta_\theta} d\theta^2 \right)
\nonumber\\
&+& \frac{(r_0^2+a^2)^2 \Delta_\theta \sin^2\theta}{\Sigma_0} \left( \frac{d\varphi}{\Xi} + \frac{2 a r_0 \, \rho \, d\tau}{(r_0^2 + a^2) \Delta_0} \right)^2,
\\
A_{[1]} &=& - \frac{Q (r_0^2 - a^2 \cos^2\theta) - 2 P a r_0 \cos\theta}{\Sigma_0 \Delta_0} \rho \, d\tau
\nonumber\\
&-& \frac{Q a r_0 \sin^2\theta - P (r_0^2 + a^2) \cos\theta + \sigma P \Sigma_0}{\Sigma_0 \Xi} d\varphi,
\nonumber
\end{eqnarray}
where
\begin{equation}
\Delta_0 = 1 \pm \frac{a^2}{L^2} \pm \frac{6 r_0^2}{L^2}, \qquad \Sigma_0 = r_0^2 + a^2 \cos^2\theta.
\end{equation}
The dual gauge potential is
\begin{eqnarray}
\bar A_{[1]} &=& - \frac{P (r_0^2 - a^2 \cos^2\theta) + 2 Q a r_0 \cos\theta}{\Sigma_0 \Delta_0} \rho \, d\tau
\nonumber\\
&-& \frac{P a r_0 \sin^2\theta + Q (r_0^2 + a^2) \cos\theta - \sigma Q \Sigma_0}{\Sigma_0 \Xi} d\varphi.
\end{eqnarray}
Note that the near horizon geometry of extremal KN-(A)dS black hole $(B = 0)$ has an enhanced symmetry with AdS$_2 \times S^2$ rather than a warped AdS$_3$~\cite{Hartman:2008pb}.

\begin{figure}[ht]
\begin{center}
\includegraphics[width=0.8\textwidth]{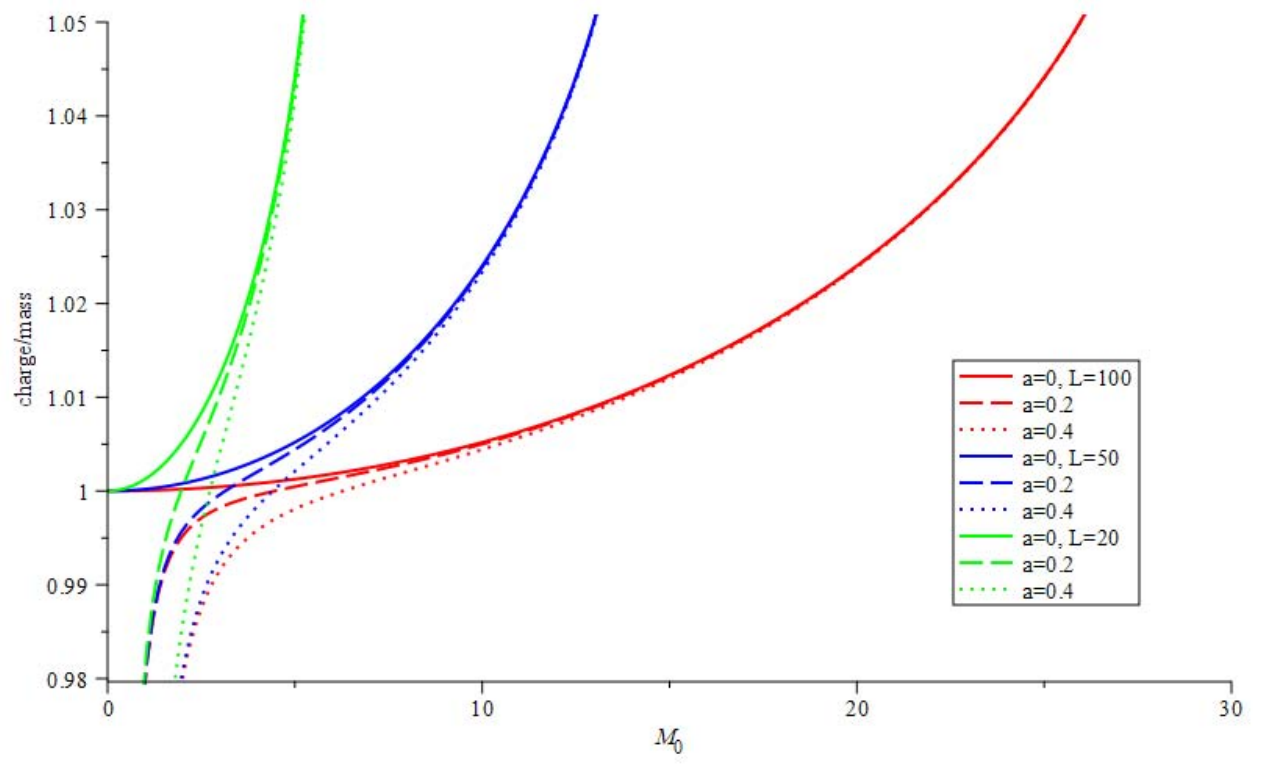}
\caption{Charge-mass ratio in dS. For a given $M_0$, the charge-mass ratio is larger for a small dS radius than for a large dS radius. The rotation $a$ lowers the charge-mass ratio for given $M_0$ and $L$.} \label{fig_QM}
\end{center}
\end{figure}

\begin{figure}[ht]
\begin{center}
\includegraphics[width=0.8\textwidth]{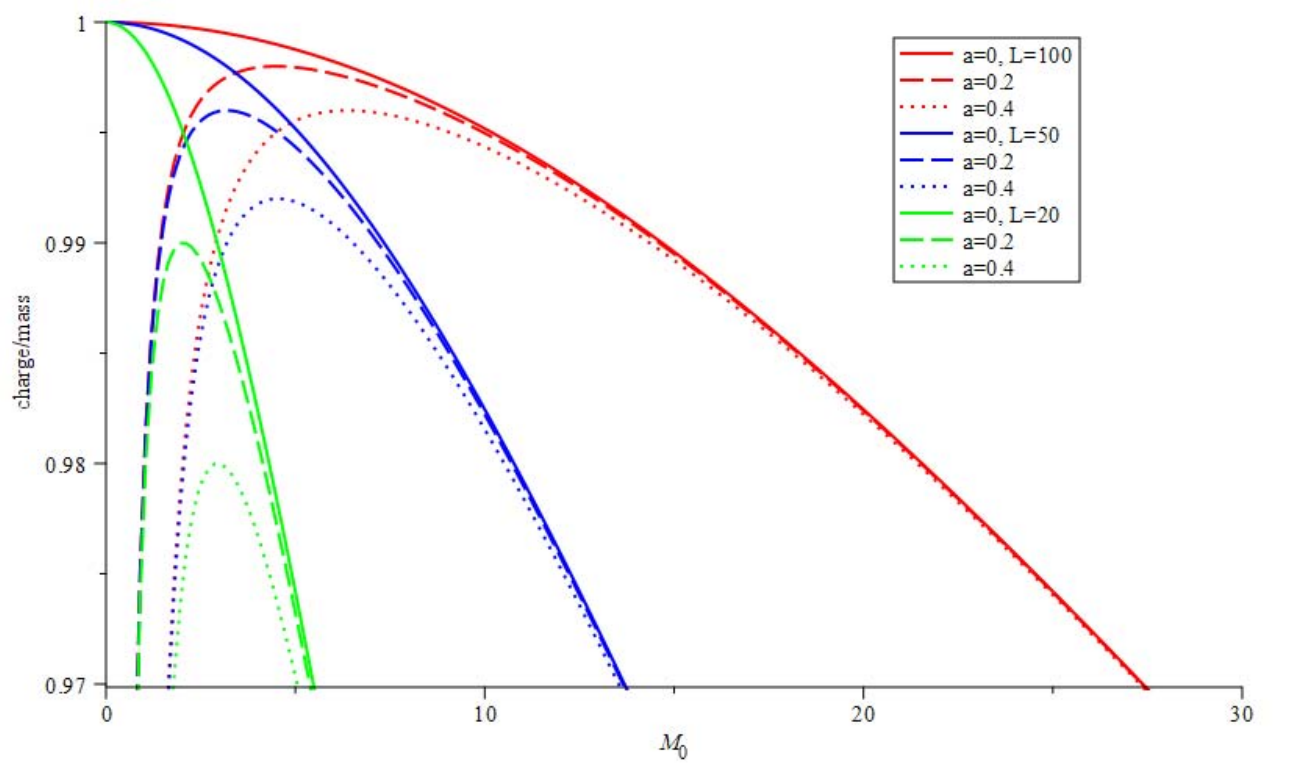}
\caption{Charge-mass ratio in AdS. For a given $M_0$, the charge-mass ratio is smaller for a small AdS radius than for large AdS radius. The angular momentum $a$ also lowers the charge-mass ratio for given $M_0$ and $L$.} \label{fig_QMa}
\end{center}
\end{figure}

The charge-mass ratio, $\sqrt{Q^2/\Xi^2 + P^2/\Xi^2}/(M_0/\Xi^2)$, in the dS space is shown in Fig.~\ref{fig_QM}. Note that there is an upper limit of $M_0$ in dS space corresponding the triple degeneracy $r_+ = r_- = r_C$, the so-called ultracold horizon. The charge-mass for AdS black holes is shown in Fig.~\ref{fig_QMa}.

\section{Schwinger Emission of Dyons}\label{secIII}
In this section, we are going to briefly summarize the computation of the production rate for the Schwinger effect occurring in the near-horizon region of near-extremal Kerr-Newman (KN) black holes in (A)dS space. The detailed calculation can be found in~\cite{Chen:2016caa, Chen:2017mnm} for the case with vanishing cosmological constant. The Klein-Grodon equation for a charged complex scalar field $\Psi$ carrying mass $m$, electric charge $q$ and magnetic charge $p$
\begin{equation}
\left( \nabla_\mu - i q A_\mu - i p \bar A_\mu \right) \left( \nabla^\mu - i q A^\mu - i p \bar A^\mu \right) \Psi - m^2 \Psi = 0,
\end{equation}
with the following ansatz,
\begin{equation}
\Psi(\tau, \rho, \theta, \varphi) = \exp\left( - i \omega \tau + i \frac{n - \sigma (q P - p Q)}{\Xi} \varphi \right) R(\rho) S(\theta),
\end{equation}
where $\omega$ is frequency, $n$ is separation integer constant from coordinate $\varphi$, and $\sigma (q P - p Q)$ term is introduced to remove the Dirac string singularity, can be separated to the angular part
{\small \begin{eqnarray}
&& \frac1{\sin\theta} \partial_\theta (\Delta_\theta \sin\theta \, \partial_\theta S)
\\
\!&-&\! \left( \frac{[n \Sigma_0 \!+\! (q Q \!+\! p P) a r_0 \sin^2\theta \!-\! (q P \!-\! p Q) (r_0^2 \!+\! a^2) \cos\theta]^2}{(r_0^2 + a^2) \Delta_\theta \sin^2\theta} \!-\! m^2 a^2 \sin^2\theta \!-\! \lambda_l \right) S \!=\! 0,
\nonumber
\end{eqnarray} }
and the radial part
{\small \begin{eqnarray}
&& \Delta_0 \partial_\rho \left( (\rho^2 - B^2) \partial_\rho R \right)
\\
&+& \left( \frac{[\omega (r_0^2 + a^2) \Delta_0 - (q Q + p P) (r_0^2 - a^2) \rho + 2 n a r_0 \rho]^2}{(r_0^2 + a^2)^2 \Delta_0 (\rho^2 - B^2)} - m^2 (r_0^2 + a^2) - \lambda_l \right) R = 0.
\nonumber
\end{eqnarray} }
In this paper we are considering the probe limit in which the back reaction of scalar field to the Einstein equation is neglected.

\begin{figure}[ht]
\begin{center}
\includegraphics[width=0.8\textwidth]{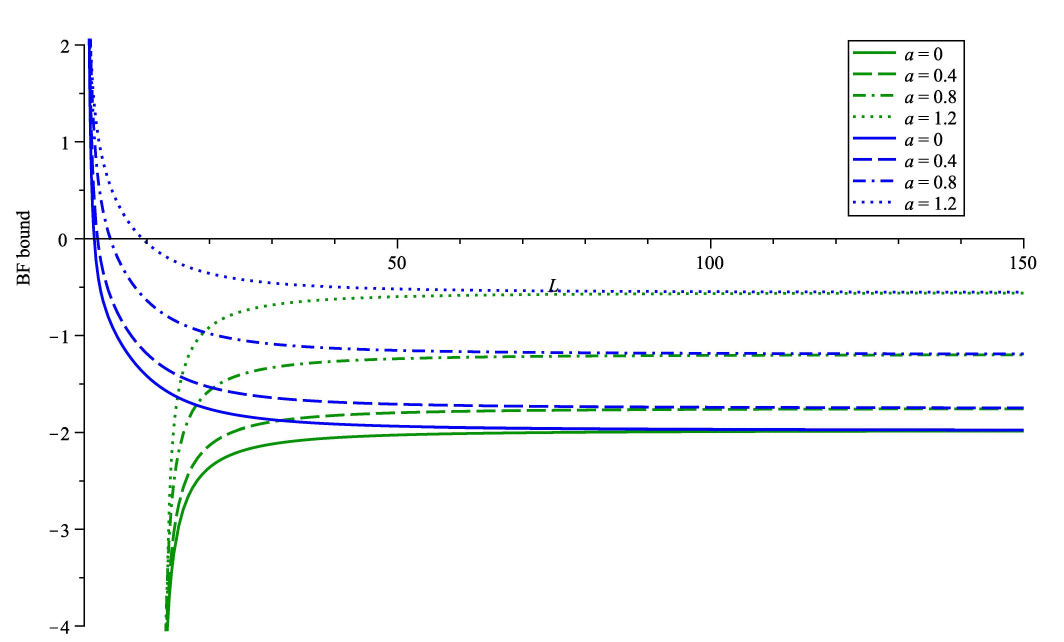}
\caption{The plot of BF bound $m_\mathrm{eff}^2 + \frac{\Delta_0}{4 (r_0^2 + a^2)}$ for AdS (blue) and dS (green) with parameters $Q^2 + P^2 = 12, q Q + p P = 6, m = 1, \lambda_l = 0, n = 0$. The BF bound can be saturated, i.e. positive region, only in AdS.} \label{fig_BF}
\end{center}
\end{figure}

The effective mass in the radial equation, proportional to the constant term in the last parentheses, is
\begin{equation}
m_\mathrm{eff}^2 = m^2 - \frac{[2 n a r_0 - (q Q + p P) (r_0^2 - a^2)]^2}{(r_0^2 + a^2)^3 \Delta_0} + \frac{\lambda_l}{r_0^2 + a^2},
\end{equation}
and the BF bound, a stability condition in the near-horizon AdS space i.e. $m_\mathrm{eff}^2 \ge -1/4 L_\mathrm{AdS}^2$, is
\begin{equation}
m_\mathrm{eff}^2 \ge -\frac{\Delta_0}{4 (r_0^2 + a^2)}.
\end{equation}
The pair production, corresponding to the unstable modes, occurs when the BF is violated~\cite{Pioline:2005pf, Kim:2008xv}. As plotted in Fig.~\ref{fig_BF}, it is clear that the pair production always occurs in dS space, but for AdS there is a threshold of $L$ which increases as rising $a$ (see Table~\ref{tab_BF}).

\begin{table}[h]
\begin{center}
\begin{tabular}{|c|c|}
\hline
\hspace{1cm} $a$ \hspace{1cm} & \hspace{1cm} $L$ \hspace{1cm}
\\ \hline\hline
$0$ & $1.5749$
\\ \hline
$0.4$ & $2.1085$
\\ \hline
$0.8$ & $4.1391$
\\ \hline
$1.2$ & $9.2893$
\\ \hline
\end{tabular}
\caption{Critical values of $a$ and $L$ of the BF bound in AdS: $Q^2 + P^2 = 12, q Q + p P = 6$, $m = 1, \lambda_l = 0, n = 0$.} \label{tab_BF}
\end{center}
\end{table}

\begin{figure}[ht]
\begin{center}
\includegraphics[width=0.8\textwidth]{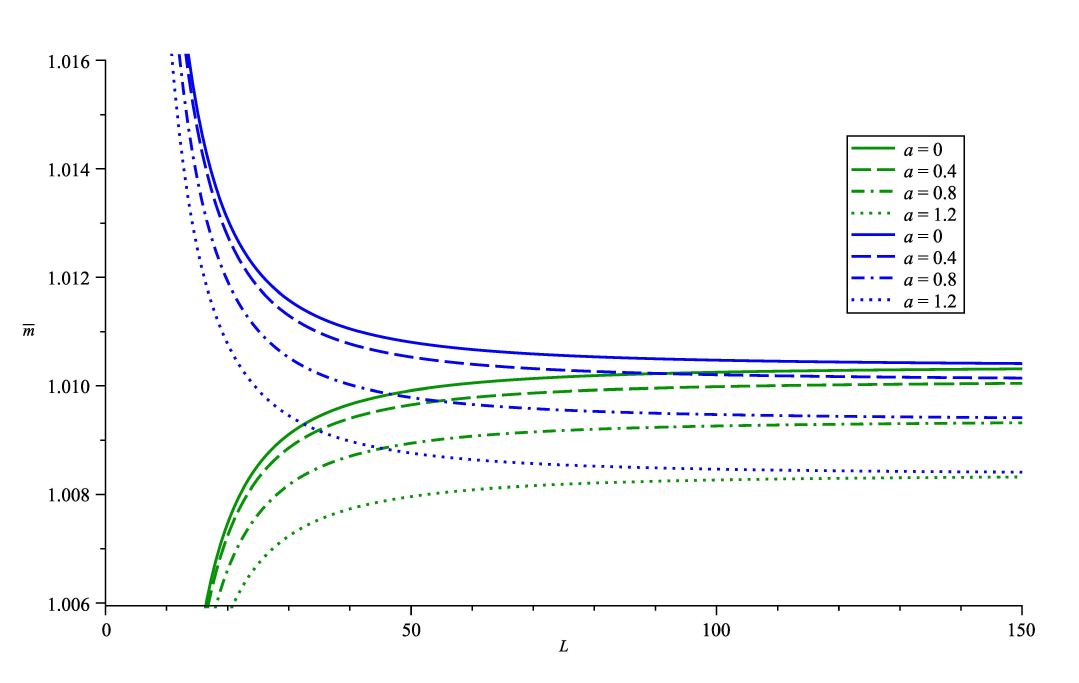}
\caption{Inertial mass $\bar m$ in AdS (blue) and dS (green) with parameters $Q^2 + P^2 = 12, q Q + p P = 6, m = 1, \lambda_l = 0, n = 0$.} \label{fig_bmL}
\end{center}
\end{figure}

The solution of radial equation is given by the hypergeometric functions~\cite{Chen:2016caa, Chen:2017mnm}
\begin{eqnarray} \label{sol}
R(\rho) &=& c_1 (\rho - B)^{\frac{i}2 (\tilde \kappa - \kappa)} (\rho + B)^{\frac{i}2 (\tilde \kappa + \kappa)}
\nonumber\\
&\times& F\left( \frac12 + i \tilde \kappa + i \mu, \frac12 + i \tilde \kappa - i \mu; 1 + i \tilde \kappa - i \kappa; \frac12 - \frac{\rho}{2B} \right)
\nonumber\\
&+& c_2 (\rho - B)^{- \frac{i}2 (\tilde \kappa - \kappa)} (\rho + B)^{\frac{i}2 (\tilde \kappa + \kappa)}
\nonumber\\
&\times& F\left( \frac12 + i \kappa + i \mu, \frac12 + i \kappa - i \mu; 1 - i \tilde \kappa + i \kappa; \frac12 - \frac{\rho}{2B} \right),
\end{eqnarray}
with the generalized parameters
\begin{eqnarray}
&& \tilde\kappa = \frac{\omega}{B}, \qquad \kappa = \frac{(q Q + p P) (r_0^2 - a^2) - 2 n a r_0}{(r_0^2 + a^2) \Delta_0},
\nonumber\\
&& \mu = \sqrt{\kappa^2 - \frac{m^2 (r_0^2 + a^2)}{\Delta_0} - \frac{\lambda_l}{\Delta_0} - \frac14}.
\end{eqnarray}
The solution will reduce to two-parameter Whittaker function for the extremal limit $B \to 0$. The violation of the BF bound leads to real value of $\mu$ which implies the existence of pair production. According the analysis in~\cite{Chen:2016caa, Chen:2017mnm}, the mean number of Schwinger effect is remarkably given by a universal formula
\begin{eqnarray}
\mathcal{N} = \Biggl( \frac{\mathrm{e}^{-2\pi (\kappa - \mu)} - \mathrm{e}^{-2\pi (\kappa + \mu)}}{1 + \mathrm{e}^{-2\pi (\kappa + \mu)}} \Biggr) \times \Biggl( \frac{1 - \mathrm{e}^{-2\pi (\tilde{\kappa} - \kappa)}}{1 + \mathrm{e}^{-2\pi (\tilde{\kappa} - \mu)}} \Biggr).
\end{eqnarray}
It should be noted that for an extremal black hole with $B = 0$ and $\tilde\kappa = \infty$, the second parenthesis deduces to unity and the first parenthesis is the mean number of pairs in the AdS$_2$~\cite{Cai:2014qba}.
The exponentials can be rewritten as the Boltzmann factors in thermodynamics with
\begin{equation}
2 \pi (\kappa - \mu) = \bar{m}/T_\mathrm{eff}, \qquad 2 \pi (\kappa + \mu) = \bar{m}/\bar{T}_\mathrm{eff},
\end{equation}
in which the associated ``inertial mass'' $\bar m$, the Unruh temperature and the effective temperatures for the Schwinger effect are defined as
\begin{eqnarray}
&& \bar m^2 = m^2 + \frac{\lambda_l}{r_0^2 + a^2} + \frac{\Delta_0}{4 (r_0^2 + a^2)}, \qquad T_\mathrm{U} = \frac{\Delta_0 \kappa}{2 \pi \bar m (r_0^2 + a^2)},
\\
&& T_\mathrm{eff} = T_\mathrm{U} + \sqrt{T_\mathrm{U}^2 - \frac{\Delta_0}{4 \pi^2 (r_0^2 + a^2)}}, \quad {\bar T}_\mathrm{eff} = T_\mathrm{U} - \sqrt{T_\mathrm{U}^2 - \frac{\Delta_0}{4 \pi^2 (r_0^2 + a^2)}}.
\nonumber
\end{eqnarray}
Note that the inertial mass increases in the AdS space while it decreases in the dS space, and the rotation reduces its value, as shown in Fig.~\ref{fig_bmL}.

\begin{figure}[ht]
\begin{center}
\includegraphics[width=0.8\textwidth]{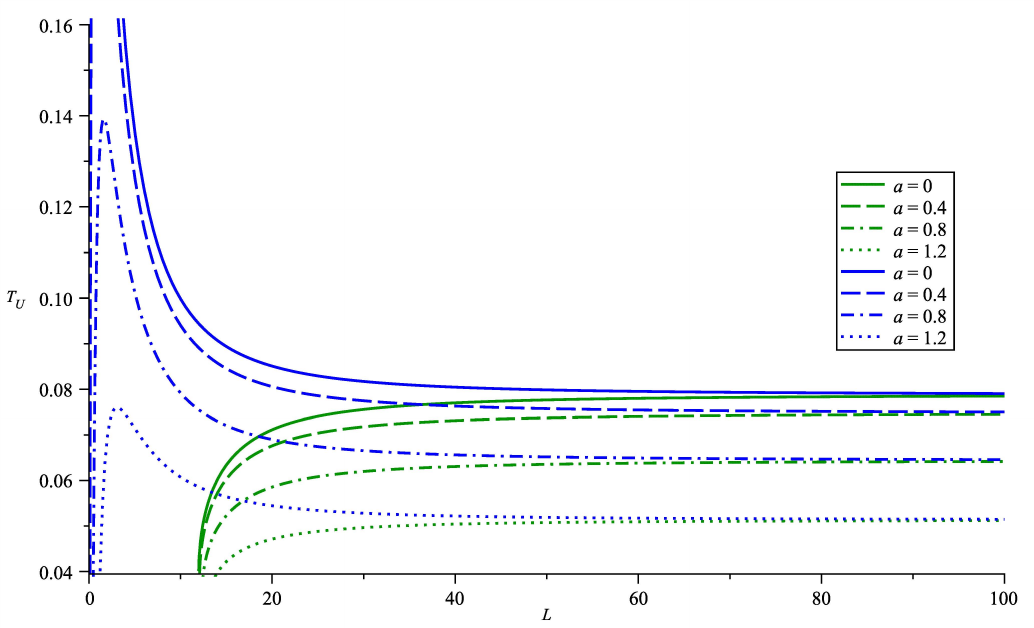}
\quad
\includegraphics[width=0.8\textwidth]{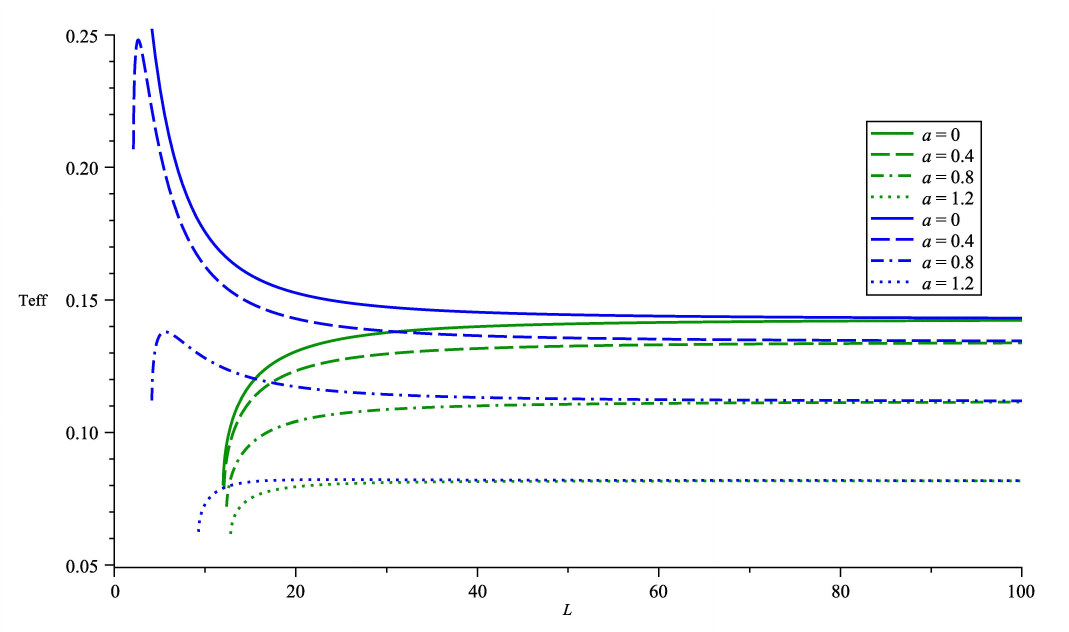}
\caption{Unruh temperature $T_\mathrm{U}$ and effective temperature $T_\mathrm{eff}$ in AdS (blue) and dS (green) with parameters $Q^2 + P^2 = 12, q Q + p P = 6, m = 1, \lambda_l = 0, n = 0$.} \label{fig_TL}
\end{center}
\end{figure}

The Unruh and effective temperatures are shown in Fig.~\ref{fig_TL}. Both temperatures basically are inversely proportional to $r_0$ since the electric field is stronger when the horizon is closer to the origin. Generally the angular momentum reduces these temperatures via enlarging the horizon. However, in AdS space, the dominant effect by $a$ or $L$ switches at almost the same value of $L$, which causes the U-turns in the Unruh temperature. Moreover, the $r_0$ can reduce to a value closed to $a$, then the factor $r_0^2 - a^2$ (assuming $n = 0$) significantly suppresses the value of $\kappa$ and, accordingly the Unruh temperature. The effective temperature is physically meaningful only in the situation that pair production occurs, i.e. violating the BF bound, otherwise it does not take real values anymore. The left endpoints of the effective temperature are just the thresholds of the BF bound.

Finally, we obtain the remarkable thermal interpretation for the Schwinger effects~\cite{Kim:2015kna}
\begin{eqnarray}
\mathcal{N} &=& \underbrace{\Biggl( \frac{\mathrm{e}^{- \bar m/T_\mathrm{eff}} - \mathrm{e}^{- \bar m/\bar T_\mathrm{eff}}}{1 + \mathrm{e}^{- \bar m/\bar T_\mathrm{eff}}} \Biggr)}_\textrm{Schwinger effect in AdS$_2$}
\nonumber\\
&\times& \Biggl\{ \mathrm{e}^{\bar m/T_\mathrm{eff}} \underbrace{\Biggl( \mathrm{e}^{- \bar m/T_\mathrm{eff}} \frac{1 - \mathrm{e}^{-(\omega + q \Phi_\mathrm{H} + p \bar\Phi_H + n \Omega_H)/T_\mathrm{H}}}{1 + \mathrm{e}^{- \bar m/T_\mathrm{eff}} \mathrm{e}^{-(\omega + q \Phi_\mathrm{H} + p \bar\Phi_H + n \Omega_H)/T_\mathrm{H}}} \Biggr)}_\textrm{Schwinger effect in Rindler$_2$} \Biggr\}.
\end{eqnarray}
The emission formula has a universal factorization of the Schwinger formula in the AdS$_2$~\cite{Cai:2014qba} and another Schwinger formula in the two-dimensional Rindler space~\cite{Gabriel:1999yz}. We may interpret the factorization such that the near-extremal black holes with non-zero $B$ have the near-horizon geometry which behaves a two-dimensional Rindler space near $\rho = B$ and gives the non-vanishing Hawking temperature though it becomes the AdS$_2$ for large $\rho$. The extremal black hole limit has the zero Hawking temperature and the second factor reduces to unity and the mean number is the Schwinger formula in the AdS$_2$.
Here, the thermodynamic quantities in the near-horizon geometry reduce to
\begin{eqnarray}
&& T_H = \frac{B}{2 \pi}, \qquad \Omega_H = \frac{2 a r_0 B}{(r_0^2 + a^2) \Delta_0}, \qquad \Phi_H = - \frac{Q (r_0^2 - a^2) B}{(r_0^2 + a^2) \Delta_0},
\nonumber\\
&& \bar\Phi_H = - \frac{P (r_0^2 - a^2) B}{(r_0^2 + a^2) \Delta_0},
\end{eqnarray}
which are the near extremal limit $r_+ = r_0 + \epsilon B$ of the corresponding quantities in~(\ref{TSOP}):
\begin{eqnarray}
&& \hat T_\mathrm{H} \to \frac{B}{2 \pi} \frac{\Delta_0 \epsilon}{r_0^2 + a^2}, \qquad \hat \Omega_\mathrm{H} \to \frac{a \Xi}{r_0^2 + a^2} - \frac{2 a r_0 B}{(r_0^2 + a^2) \Delta_0} \frac{\Xi \Delta_0 \epsilon}{r_0^2 + a^2},
\nonumber\\
&& \hat\Phi_\mathrm{H} \to \frac{Q r_0}{r_0^2 + a^2} - \frac{Q (r_0^2 - a^2) B}{(r_0^2 + a^2) \Delta_0} \frac{\epsilon \Delta_0}{r_0^2 + a^2}.
\end{eqnarray}
The limit is as follows: the first term in $\hat \Omega_\mathrm{H}$ is eliminated by transformation $\varphi \to \varphi + \hat\Omega_H t$, and the first term in $\hat \Phi_\mathrm{H}$ is discarded when taking the near horizon limit (it is a divergent constant). The factor $\frac{\Delta_0 \epsilon}{r_0^2 + a^2}$ is simply due to the rescaling of time, and $\Xi$ in $\hat \Omega_\mathrm{H}$ is from the $\varphi$ dependence in the ansatz of the scalar field.

\section{Conclusion}\label{secIV}
We have studied the emission of electric and/or magnetic charges from the near-extremal dyonic KN black holes in the (A)dS space. The dyonic KN black holes in the (A)dS space have the limits: (i) the dyonic KN black holes in the asymptotically flat space when the (A)dS radius is infinite, (ii) the dyonic RN black holes in the (A)dS space when the rotation of the hole vanishes and (iii) the KN black holes in the (A)dS space when the magnetic charge is zero or the magnetic KN black holes in the (A)dS space when the electric charge is zero. Using the AdS geometry of near-horizon region, we have found the solutions of the dyons in terms of the hypergeometric function and obtained the Schwinger effect by computing the mean number of the emitted dyons.

We have shown that the emission formula still has a universal factorization of the Schwinger formula in the AdS$_2$ and another Schwinger formula in the two-dimensional Rindler space determined by the effective temperature and the Hawking temperature with the chemical potentials of electric and magnetic charges and angular momentum. The emission formula recovers the formulae for RN or KN black holes in the asymptotically flat space and also the formula for RN black holes in the (A)dS space. Further, we have shown that there exists a parameter region, known as the Breitenloher-Freedman bound for the stability of dyonic KN black holes in the AdS space against both the emission of charges and Hawking radiation.

We have observed that the Schwinger effect is affected by the (A)dS boundary and the rotation of the black hole as follows:
\begin{itemize}

\item The boundary AdS space compresses the horizon radius while the dS space stretches the radius, which in turn strengthens or weakens the electric field on the horizon where the pair are mostly produced. This holds for RN black holes or KN black holes in the boundary (A)dS space. Hence, the Schwinger effect for a fixed angular momentum is enhanced in AdS space and suppressed in dS space.

\item The rotation of near-extremal KN black hole decreases the ``inertial mass'' of a charge since the dragging of particles around the event horizon lightens the effective mass along the radial direction, but increases the horizon radius and reduces the electric field on the horizon. For a large radius of the (A)dS space, the decreasing electric field affects more than the decreasing inertial mass in the charge acceleration for the Unruh temperature. Hence, for a large radius of the (A)dS space, the angular momentum decreases the Unruh and effective temperatures and thus, suppresses the Schwinger effect. However, for a small radius of the AdS boundary, as the angular momentum changes, there is a crossover of the horizon radius at almost the same value of the AdS radius: below the crossover the horizon radius decreases as the angular momentum while above the crossover the horizon radius increases as the angular momentum. The competitive effect of the crossover of the horizon radius and the decreasing inertial mass results in the convexity of the Unruh and effective temperatures.

\item In the AdS space the KN black hole has a BF bound that depends on the angular momentum and provides a threshold for the Schwinger effect in AdS space. The radius of the AdS space for the BF bound increases as the angular momentum increases. In the dS space the charge to mass ratio has an upper bound due to the triple degeneracy of horizons: the Cauchy horizon, black hole horizon and cosmological horizon. The ratio decreases as the angular momentum increases for small mass of  the black hole but approaches the same value until the upper bound.

\end{itemize}

\acknowledgments
The work of C.M.C. was supported by the National Science and Technology Council of the R.O.C. (Taiwan) under the grants MOST 109-2112-M-008-010 and 110-2112-M-008-009.
The work of S.P.K. was supported in part by National Research Foundation of Korea (NRF) funded by the Ministry of Education (2019R1I1A3A01063183).


\begin{thebibliography}{99}

\bibitem{Griffiths:2012}
  J.~B.~Griffiths and J.~Podolsk{\'y},
  \textit{Exact Space-Times in Einstein's General Relativity},
  (Cambridge University Press, Cambridge, 2012).

\bibitem{Kunduri:2008rs}
  H.~K.~Kunduri and J.~Lucietti,
  ``A Classification of near-horizon geometries of extremal vacuum black holes,''
  J. Math. Phys. \textbf{50}, 082502 (2009)
  [arXiv:0806.2051 [hep-th]].

\bibitem{Chen:2016caa}
  C.-M.~Chen, S.~P.~Kim, J.-R.~Sun and F.-Y.~Tang,
  ``Pair Production in Near Extremal Kerr-Newman Black Holes,''
  Phys.\ Rev.\ D {\bf 95}, 044043 (2017)
  [arXiv:1607.02610 [hep-th]].

\bibitem{Chen:2020mqs}
  C.-M.~Chen and S.~P.~Kim,
  ``Schwinger Effect from Near-extremal Black Holes in (A)dS Space,''
  Phys. Rev. D \textbf{101} (2020) no.8, 085014
  [arXiv:2002.00394 [hep-th]].

\bibitem{Montero:2019ekk}
  M.~Montero, T.~Van Riet and G.~Venken,
  ``Festina Lente: EFT Constraints from Charged Black Hole Evaporation in de Sitter,''
  JHEP \textbf{01}, 039 (2020)
  [arXiv:1910.01648 [hep-th]].


\bibitem{Hawking:1974sw}
  S.~W.~Hawking,
  ``Particle Creation by Black Holes,''
  Commun.\ Math.\ Phys.\  {\bf 43}, 199 (1975)
  Erratum: [{\it Commun.\ Math.\ Phys.}\  {\bf 46}, 206 (1976)].

\bibitem{Sauter:1932ab}
  F.~Sauter,
  ``Zum Kleinschen Paradoxon,''
  Z.\ Phys.\ {\bf 73}, 547 (1932).

\bibitem{Schwinger:1951nm}
  J.~S.~Schwinger,
  ``On gauge invariance and vacuum polarization,''
  Phys.\ Rev.\  {\bf 82}, 664 (1951).

\bibitem{Belgiorno:2007va}
  F.~Belgiorno and S.~L.~Cacciatori,
  ``Quantum Effects for the Dirac Field in Reissner-Nordstrom-AdS Black Hole Background,''
  Class.\ Quant.\ Grav.\  {\bf 25}, 105013 (2008)
  [arXiv:0710.2014 [gr-qc]].

\bibitem{Belgiorno:2008mx}
  F.~Belgiorno and S.~L.~Cacciatori,
  ``Massive Dirac particles on the background of charged de-Sitter black hole manifolds,''
  Phys.\ Rev.\  D {\bf 79}, 124024 (2009)
  [arXiv:0810.1642 [gr-qc]].


\bibitem{Chen:2012zn}
  C.-M.~Chen, S.~P.~Kim, I.-C.~Lin, J.-R.~Sun and M.-F.~Wu,
  ``Spontaneous Pair Production in Reissner-Nordstrom Black Holes,''
  Phys.\ Rev.\ D {\bf 85}, 124041 (2012)
  [arXiv:1202.3224 [hep-th]].

\bibitem{Chen:2014yfa}
  C.-M.~Chen, J.-R.~Sun, F.-Y.~Tang and P.-Y.~Tsai,
  ``Spinor particle creation in near extremal Reissner–Nordström black holes,''
  Class.\ Quant.\ Grav.\  {\bf 32}, 195003 (2015)
  [arXiv:1412.6876 [hep-th]].


\bibitem{Chen:2017mnm}
  C.-M.~Chen, S.~P.~Kim, J.-R.~Sun and F.-Y.~Tang,
  ``Pair production of scalar dyons in Kerr–Newman black holes,''
  Phys.\ Lett.\ B {\bf 781}, 129 (2018)
  [arXiv:1705.10629 [hep-th]].


  \bibitem{Zhang:2020apg}
  J.~Zhang, Y.-Y.~Lin, H.-C.~Liang, K.-J.~Chi, C.-M.~Chen, S.~P.~Kim and J.-R.~Sun,
  ``Pair production in Reissner-Nordstr\"om-Anti de Sitter black holes,''
  Chin. Phys. C \textbf{45} (2021) no.6, 065105
  [arXiv:2003.06398 [hep-th]].

  \bibitem{Cai:2020trh}
  R.-G.~Cai, C.-M.~Chen, S.~P.~Kim and J.-R.~Sun,
  ``Schwinger effect in near-extremal charged black holes in high dimensions,''
  Phys. Rev. D \textbf{101} (2020) no.10, 105015
  [arXiv:2004.00735 [hep-th]].

\bibitem{Dickson:1914}
L.~E.~Dickson, \textit{Elementary Theory of Equations}, (John Wiley \& Sons Inc., New York, 1914).

\bibitem{Hartman:2008pb}
  T.~Hartman, K.~Murata, T.~Nishioka and A.~Strominger,
  ``CFT Duals for Extreme Black Holes,''
  JHEP \textbf{04}, 019 (2009)
  [arXiv:0811.4393 [hep-th]].

\bibitem{Pioline:2005pf}
  B.~Pioline and J.~Troost,
  ``Schwinger pair production in AdS(2),''
  JHEP {\bf 0503}, 043 (2005)
  [hep-th/0501169].

\bibitem{Kim:2008xv}
  S.~P.~Kim and D.~N.~Page,
  ``Schwinger Pair Production in dS(2) and AdS(2),''
  Phys.\ Rev.\  D {\bf 78}, 103517 (2008)
  [arXiv:0803.2555 [hep-th]].

\bibitem{Cai:2014qba}
  R.-G.~Cai and S.~P.~Kim,
  ``One-Loop Effective Action and Schwinger Effect in (Anti-) de Sitter Space,''
  JHEP {\bf 1409}, 072 (2014)
  [arXiv:1407.4569 [hep-th]].

\bibitem{Kim:2015kna}
  S.~P.~Kim, H.~K.~Lee and Y.~Yoon,
  ``Thermal Interpretation of Schwinger Effect in Near-Extremal RN Black Hole,''
  Int.\ J.\ Mod.\ Phys.\ D {\bf 28}, 1950139 (2019)
  [arXiv:1503.00218 [hep-th]].

\bibitem{Gabriel:1999yz}
  C.~Gabriel and P.~Spindel,
  ``Quantum charged fields in Rindler space,''
  Annals Phys.\  {\bf 284}, 263 (2000)
  [arXiv:gr-qc/9912016].

\end{thebibliography}
\end{document}